\documentclass{article}
\expandafter\let\csname equation*\endcsname\relax
\expandafter\let\csname endequation*\endcsname\relax
\usepackage{amsmath}
\usepackage{graphicx}
\usepackage{subfig}
\usepackage{algpseudocode}
\usepackage{algorithm}
\usepackage{lscape}
\usepackage[modulo]{lineno}
\usepackage{pgfplots}
\usepackage{array}
\usepackage{lscape}
\usepackage{rotating}
\usepackage{tikz}
\usepackage{varwidth}
\usepackage{rotating}
\usepackage{color}
\usetikzlibrary{decorations.pathreplacing}
\usetikzlibrary{shapes}
\usepackage[affil-it]{authblk}
\usepackage{geometry}
\usepackage[round]{natbib}
\usepackage{hyperref}
\hypersetup{
	colorlinks=true,
	citecolor=black,
	urlcolor=blue,
	linkcolor=black,
	pdfborder={0 0 0},
}

\begin{document}
	\title{Fast approximate delivery of fluence maps: the VMAT case}
	\author{Marleen Balvert$^1$, David Craft$^2$}
	\affil{$^1$ Department of Econometrics and Operations Research/Center for Economic Research (CentER), Tilburg University, PO Box 90153, 5000 LE Tilburg, The Netherlands}
	\affil{$^2$ Department of Radiation Oncology, Massachusetts General Hospital and Harvard Medical School, Boston, MA 02114, USA}
	\date{}

\maketitle
	
	\begin{abstract}
		In this article we provide a method to generate the trade-off between delivery time and fluence map matching quality for volumetric modulated arc therapy (VMAT). At the heart of our method lies a mathematical programming model that, for a given duration of delivery, optimizes leaf trajectories and dose rates such that the desired fluence map is reproduced as well as possible. This model was presented for the single map case in a companion paper \citep{dm1}. The resulting large-scale, non-convex optimization problem was solved using a heuristic approach. The single-map approach cannot directly be applied to the full arc case due to the large increase in model size, the issue of allocating delivery times to each of the arc segments, and the fact that the ending leaf positions for one map will be the starting leaf positions for the next map. In this article the method proposed in \cite{dm1} is extended to solve the full map treatment planning problem. We test our method using a prostate case and a head and neck case, and present the resulting trade-off curves. Analysis of the leaf trajectories reveal that short time plans have larger leaf openings in general than longer delivery time plans. Our method allows one to explore the continuum of possibilities between coarse, large segment plans characteristic of direct aperture approaches and narrow field plans produced by sliding window approaches. Exposing this trade off will allow for an informed choice between plan quality and solution time. Further research is required to speed up the optimization process to make this method clinically implementable.
	\end{abstract}
	
	\section{Introduction}
	Volumetric modulated arc therapy (VMAT) delivers radiation while continuously moving the gantry around the patient and simultaneously adjusting the positions of the multi-leaf collimator's leaves. This allows for a faster delivery than intensity modulated radiotherapy (IMRT) without compromising the plan quality (see, e.g., \cite{Teoh:2011} and references therein). Even faster deliveries can be achieved for both IMRT and VMAT when accepting sub-optimal treatment plans. The trade-off between treatment duration and plan quality thus lies at the heart of IMRT and VMAT treatment planning \citep{Unkelbach:2015}. In the first paper of this set of two, we have developed a model to construct the treatment time versus plan quality trade-off for the single fluence map case that applies to IMRT \citep{dm1}. The model is extended in this second paper, where the aim is to develop a method to construct the trade-off curve for a full arc VMAT treatment and allow the planner to balance treatment time and plan quality.
	
	Treatment planning for VMAT involves optimization of the leaf trajectories and dose rates. A common approach is based on treatment planning for IMRT via fluence maps. First, the 360$^{\circ}$ arc is segmented and fluence maps are optimized for each of the arc segments based on planning goals. Next, leaf trajectories and dose rates are optimized such that the fluence maps are reproduced with maximum accuracy, which is often referred to as arc sequencing. Fluence map optimization for IMRT is a well studied problem, and optimization tools to explore the Pareto frontier for various planning objectives have been developed and shown to yield good results \citep{Kierkels:2015}. This leaves the arc sequencing step to be developed specifically for VMAT. A sliding window approach, where both leaves cross the whole field, is known to be capable of replicating the fluence map to high accuracy \citep{Stein:94}. Many arc sequencing methods are based on or make use of a sliding window approach \citep{Luan:2008,vmerge}. For an extensive review of VMAT treatment plan optimization methods, see \cite{Unkelbach:2015}.
	
	The sliding window delivery technique generally yields lengthy treatments. Limiting the delivery time inevitably results in a reduction in fluence map resemblance, though the actual trade-off has not been investigated before. Our goal is to construct the plan quality versus delivery time Pareto frontier for VMAT, where plan quality is measured as the accuracy level at which the optimal fluence map is replicated.
	
	In \cite{dm1} we have proposed a method to construct the time versus plan quality trade-off curve for a single fluence map. For a given delivery time, the developed model determines the dose rates and leaf positions at discrete time points such that the difference between the optimized and the original fluence map is minimized. Solving this model for various delivery times gives the desired trade-off curve.
	
	The single fluence map model is adequate for IMRT treatment plan optimization. For VMAT however, optimizing each of the fluence maps separately does not give a feasible treatment plan due to discrepancies of the leaf positions at the transition between two consecutive maps. In this work, we extend the method from \cite{dm1} to the full arc case, so that one can generate a set of treatment plans, each of which corresponds to a different trade-off between fluence map matching and delivery time. 
	
	\section{Methods}
	We quantify the similarity between a desired and a planned fluence map as the sum of the squared differences (ssdif) of the fluence in each bixel. The treatment plan optimization model aims at finding a treatment plan (i.e., leaf trajectories and dose rates) that minimizes ssdif summed over all fluence maps for a given delivery time, while satisfying hardware constraints such as maximum leaf speed and maximum dose rate. 
	
	As a starting point, we use the optimization model for the single map case developed in \cite{dm1}, where leaf trajectories and dose rates are optimized given the amount of time that is spent on delivering this map. For clarity, we briefly restate the main ingredients of this model. Time is discretized into steps $1,...,T$, and the leaf positions of the left and the right leaf of row $i$ at time step $t$ are denoted by $L_t^i$ and $R_t^i$. Bixels are indexed $1,...,B$, and the left and right boundaries of bixel $j$ coincide with positions $j$ and $j+1$ on the physical ruler, respectively. For example, when $L_t^i=1$, the left leaf is at the left border of the first bixel at time $t$, and $L_t^i=3.5$ means that the left leaf is in the middle of bixel 3. 
	
	Fundamental to the optimization model is the exposure function, which calculates the fraction of bixel $j$ in row $i$ that is exposed at each time step \citep{dm1}:
	\begin{align}
		e(L_t^i,R_t^i,j) &= 
		\begin{cases}
			1  & \text{if } L_t^i\leq j \text{ and } R_t^i \geq j+1\\
			j+1-L_t^i  & \text{if } j< L_t^i < j+1 \text{ and } R_t^i \geq j+1\\
			R_t^i-j & \text{if }  L_t^i \leq j \text{ and } j < R_t^i < j+1\\
			R_t^i-L_t^i & \text{if }  j < L_t^i < j+1 \text{ and } j < R_t^i < j+1\\
			0 & \text{if }   L_t^i \geq j+1 \text{ or }  R_t^i \leq j.
		\end{cases} \label{eq:exposure}
	\end{align}
	In the first case, the bixel is fully exposed, while it is partially blocked by the left, the right and by both leaves in the second, third and fourth case, respectively. In the last case, the bixel is fully blocked by one of the leaves. Note that this function is piece-wise linear and non-convex. 
	
	The amount of fluence delivered to bixel $j$ in row $i$ at time step $t$ equals the exposure multiplied by the fluence rate per time step. The total fluence in this bixel for some fluence map, $g_{ij}$, is the sum of the delivered fluence over all time steps the gantry spends in the corresponding arc segment:
	\begin{equation}
		g_{ij}=\sum_{t=1}^T e(L_t^i,R_t^i,j)D_t\Delta, \label{eq:fluenceInij}
	\end{equation}
	where $D_t$ is the dose rate in MU/s at time step $t$, and $\Delta$ is the length of a time step in s.
	
	The single map model minimizes the sum of the squared differences (ssdif) between the planned and the desired fluence in each bixel, subject to linear hardware constraints such as the maximum leaf speed. As the planned fluence in each bixel is given by \eqref{eq:fluenceInij}, the optimization problem has a highly non-convex objective function \citep{dm1}. The model is therefore solved by local search from 125 random starting solutions (the local search is done using an interior point method).  
	
	The single map optimization model can be extended to the full arc case by including multiple fluence maps and replacing the objective of minimising ssdif by minimising the sum of ssdif over all fluence maps. However, this raises several issues:
	\begin{enumerate}
		\item The model requires the amount of time spent on each arc segment as an input, whereas only the total delivery time for the full arc is known.
		\item Leaf trajectories for each map must align so that the end leaf positions for one map become the start positions for the subsequent map. 
		\item In VMAT, fluence maps are delivered over an arc sector, and thus there is the potential for dosimetric degradation of the plan if the fluence maps are optimized for a single angle and delivered over a large angular sector. We limit our angular sectors to 10$^{\circ}$ to avoid this problem \citep{vmerge}. 
		\item The increase in size makes an already difficult non-convex problem even more computationally demanding.
	\end{enumerate}
	
	To solve the aforementioned issues we decompose the problem into four steps. First, the trade-off curve between delivery time and plan quality for each arc segment separately is constructed by solving the treatment plan optimization model for the single fluence map case for various delivery times. These curves are used in the second step to allocate the total delivery time for the full arc to the arc segments, which implicitly determines the gantry rotation speed. In the third step the dose rates are fixed to those that were optimal in the segment-specific optimization, and leaf trajectories are optimized for the full arc. Finally, the solution obtained in step 3 is used as a starting point for alternatingly optimizing the dose rates and the leaf trajectories to refine our solution. Each of these steps will be further explained in the remainder of this section.
	
	\bigskip
	
	\noindent {\bf Step 1: treatment plan optimization for a single arc segment}
	
	\noindent In the first step, the trade-off between plan quality and delivery time for each arc segment with its corresponding fluence map is constructed using the single map model. Details on the optimization procedure can be found in \cite{dm1}.

	\bigskip
	
	\noindent{\bf Step 2: delivery time allocation}
	
	\noindent The trade-off curves for the individual arc segments obtained in step 1 indicate how complicated the replication of each of the fluence maps is. We use this information to allocate the total delivery time over the arc segments in such a way that, if we were able to optimize for each of the fluence maps individually, the ssdif summed over all arcs is minimized. For this we solve the following small integer programming problem:
	
	\begin{align}
		\min \quad & \sum_{m \in \mathcal{M}, t \in \mathcal{T}} v_{mt} x_{mt} \label{model:timeAllocation}\\
		\text{s.t.} \quad & \sum_{m \in \mathcal{M}, t \in \mathcal{T}} tx_{mt} \leq TT \nonumber \\
		& \sum_{t \in \mathcal{T}} x_{mt} = 1 & \forall m \in \mathcal{M} \nonumber \\
		& x_{mt} \in \{0,1\}. \nonumber
	\end{align}
	Here, $x_{mt}$ is a binary variable that is equal to 1 if $t$ units of time are assigned to arc segment $m$, and 0 otherwise. Furthermore, $v_{mt}$ is the minimum ssdif for individual optimization of arc segment $m$ when its delivery time is $t$, which is known from step 1. $\mathcal{M}$ is the set of arc segments, and $\mathcal{T}$ is the set of delivery times for which the single arc segment model was solved in step 1. The total delivery time for the full arc is limited by $TT$ in the first constraint, where the parameter $t$ is equal to the subscript $t$ of $x_{mt}$. Note that the optimal value of \ref{model:timeAllocation} is a lower bound for the true ssdif (the real solution will be worse due to leaf position matching constraints).
	\bigskip
	
	\noindent {\bf Step 3: treatment plan optimization for the full arc}
	
	\noindent With the time allocation obtained in step 2, we can solve a full-arc version of the model from \cite{dm1} in the third step. As this model optimizes the trajectories for the whole arc at once, this automatically ensures a smooth transition of leaf positions between the arc segments. However, as noted before, the resulting model is too large to solve. A natural way to decompose the problem is to solve it for each arc segment separately and fix the start and end positions of the leaves to allow for a smooth transition from one arc segment to the next, as proposed in \cite{coupled}. Fixing the leaf start and end positions imposes restrictions on what trajectory types can be used, leaving little freedom for the optimization problem. We have tested several decision rules for fixing the leaf positions at the transition points, but none of them produced treatment plans that yield fluence maps similar to the desired ones. 
	
	Instead of decomposing the problem into separate optimizations for each arc segment, one could decouple the rows and still optimize the trajectories of a single leaf pair for the full arc at once. Although the leaf trajectories of the leaf pairs are mechanically independent (assuming that interdigitation is allowed), the dose rate variable couples the individual rows since the dose rate applies to the entire field.  Therefore, in order to allow for separate optimization of the leaf trajectories, we fix the dose rates to those that were optimal in the arc segment-specific optimization.
	
	In the following, since we consider a single row, we drop the leaf row index $i$. Denote the desired fluence in map $m$ for bixel $j$ by $f_{mj}$, and the fluence obtained from the treatment plan by $g_{mj}$, $m=1,...,M$ and $j=1,...,B$.  Using function \eqref{eq:fluenceInij}, the sum of squared differences between the desired and planned fluences can be minimized using the following optimization problem for a single row:
	\begin{align}
		\text{min } \quad & \sum_{m=1}^M \sum_{j=1}^B (f_{mj} - g_{mj})^2 \nonumber  \\
		\text{s.t.} \quad & g_{mj} = \sum_{t=s_m}^{e_m} e(L_t,R_t,j)D_t \Delta \nonumber \\
		& L_t \leq R_t & \forall t \label{model:hw1} \\
		& L_t - c \leq L_{t+1} \leq L_t + c & \forall t=1\ldots T-1 \label{model:hw2} \\
		& R_t - c \leq R_{t+1} \leq R_t + c & \forall t=1\ldots T-1 \label{model:hw3} \\
		& L_t \geq 1 & \forall t \label{model:hw4} \\
		& R_t \leq B+1 & \forall t. \label{model:hw5}
	\end{align}
	where $s_m$ and $e_m$ are the first and last time step used to deliver map $m$ (note that these come from the solution to formulation (\ref{model:timeAllocation})). The hardware constraints ensure that the left leaf is always to the left of the right leaf (constraint \eqref{model:hw1}), and that the leaves move at a speed of at most $c$ bixels per time step (constraints \eqref{model:hw2} and \eqref{model:hw3}). The leaves cannot move beyond the first and the last bixel, so $L_t, R_t \in [1,B+1]$  (constraints \eqref{model:hw4} and \eqref{model:hw5}).
	
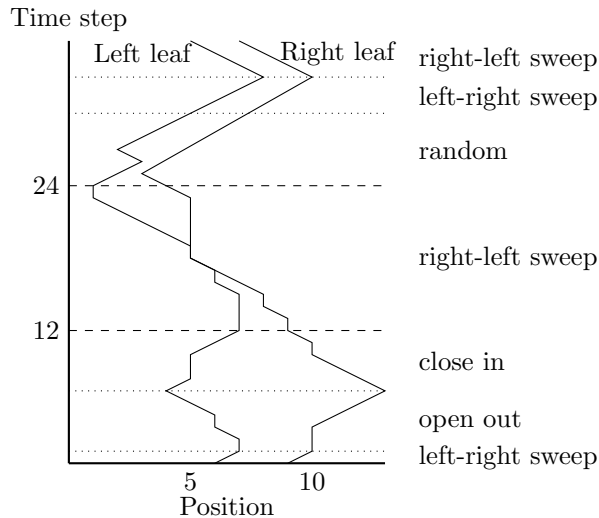
\begin{figure}[!htbp]
		\centering
		\begin{tikzpicture}[xscale=0.8,yscale=0.8,x=0.4cm,y=0.2cm]
		\draw[thick] (0,1)--(13,1);
		\draw[thick] (0,1)--(0,36);
		\draw (6.5,-1) node[anchor=north] {Position};
		\draw (0,36) node[anchor=south] {Time step};
		\draw (5.4,35) node[anchor=east] {Left leaf};
		\draw (8.3,35) node[anchor=west] {Right leaf};
		\draw(0,12) node[anchor=east] {12};
		\draw(0,24) node[anchor=east] {24};
		\draw (5,1) node[anchor=north] {5};
		\draw (10,1) node[anchor=north] {10};
		%left leaf trajectory
		\draw (6,1)--(7,2)--(7,3)--(6,4)--(6,5)--(4,7)--(5,8)--(5,10)--(7,12)--(7,15)--(6,16)--(6,17)--(5,18)--(5,19)--(1,23)--(1,24)--(3,26)--(2,27)--(8,33)--(5,36);
		%right leaf trajectory
		\draw (9,1)--(10,2)--(10,4)--(13,7)--(10,10)--(10,11)--(9,12)--(9,13)--(8,14)--(8,15)--(5,18)--(5,23)--(3,25)--(10,33)--(7,36);
		\draw[dashed] (0,12)--(13,12);
		\draw[dashed] (0,24)--(13,24);
		\draw[dotted] (0,2)--(13,2);
		\draw[dotted] (0,7)--(13,7);
		\draw[dotted] (0,30)--(13,30);
		\draw[dotted] (0,33)--(13,33);
		\draw (14,1.5) node[anchor=west] {left-right sweep};
		\draw (14,4.5) node[anchor=west] {open out};
		\draw (14,9.5) node[anchor=west] {close in};
		\draw (14,18) node[anchor=west] {right-left sweep};
		\draw (14,27) node[anchor=west] {random};
		\draw (14,31.25) node[anchor=west] {left-right sweep};
		\draw (14,34.5) node[anchor=west] {right-left sweep};
		\end{tikzpicture}
		\caption{Example of a random leaf trajectory (solid lines) for three arc segments with 12 time steps allocated to each of them. The arc segments are indicated by the dashed lines. The arc segments are subdivided into three, one and three subsegments, respectively, (dotted lines). The leaf trajectories of these seven subsegments are a left-right sweep, an open-out trajectory, a close-in trajectory, a left-right sweep a random trajectory, a left-right sweep and a right-left sweep, respectively.}\label{fig:trajExample}
	\end{figure}

	The many local minima this highly non-convex model has can be identified by solving the model using an interior-point method for various starting solutions. For the single map case, starting solutions were generated by randomly selecting a trajectory type (left-right, right-left, close-in, open-out or random) and generating a trajectory of this type (for details, see \cite{dm1}). We could use the same approach for the full arc case by selecting a random type for each of the maps and generate a trajectory accordingly. The main drawback of this approach is a lack of freedom: if the first map has a left-right trajectory, the leaves end up all the way to the right and the next trajectory can only be a right-left sweep or a random trajectory. In order to allow for more flexibility, each of the arc segments is subdivided into one (so not subdivided), two or three of subsegments of random length, for which a trajectory is generated according to a randomly selected trajectory type. The number of subsegments is randomly selected as well. An example of random leaf trajectories with three arc segments is given in Figure \ref{fig:trajExample}.
	\bigskip
	
	\noindent {\bf Step 4: alternate improvement of dose rate schedule and leaf trajectories}
	
	\noindent The treatment plans obtained in step 3 have leaf trajectories that are optimized with respect to a given dose rate schedule. However, the current dose rate schedule is not necessarily optimal for the current leaf trajectories. Therefore, we re-optimize the dose rates by fixing the leaf trajectories, for which a quadratic programming problem needs to be solved. As the current leaf trajectories may not be optimal for the new dose rate schedule, we then re-optimize the leaf trajectories using the approach in step 3, though with only one start solution, namely the current leaf trajectories. This process is repeated until the improvement per iteration drops below 1\%.
	
	\section{Results}
	Our method is tested using the same cases as in \cite{dm1}: a prostate case and a head-and-neck case, both from the publically available CORT dataset \citep{cortdoi}. We assume a maximum leaf speed of 3 cm/s and a maximum dose rate of 10 MU/s. The bixel width is 1 cm for the prostate case and 0.5 cm for the head and neck case, and the arc is divided into 36 arc segments for both cases. Fluence maps are obtained by solving a dosimetric convex optimization, see e.g. \citep{chen2012including} (the fluence map optimization problem has been described many times; so as to not distract from the fluence map matching problem, we intentionally do not discuss this problem). Time steps of 1/3 s are used for the prostate case. Since the fluence maps for the head and neck case are larger and thus result in a larger model, time step size was chosen to be 1/2 s. The model was solved using the fmincon function in Matlab Release R2014a (The Mathworks, Inc., Natick, USA). 
	
	Leaf trajectories and dose rates are optimized for total delivery times varying from 60 to 180 seconds with step sizes of 15 seconds for the prostate case, and from 75 to 345 seconds with step sizes of 90 seconds for the head and neck case. For both cases, we show the time versus fluence map resemblance (ssdif) trade-off curves (Figures \ref{fig:prosTradeOff} and \ref{fig:HNTradeOff}). The dotted line shows the total ssdif if the treatment plans for each arc segment were optimized individually, which coincides with the optimal objective values of the time allocation problem in step 2. This is a lower bound for the actual full arc trade-off, as generating a full arc treatment plan is more restrictive than optimizing the arc segments individually. The solid and dashed lines show the results for the full arc treatment plan optimization using optimized dose rate schedules and constant maximum dose rates (i.e. all dose rates equal to 10 MU/s), respectively. For these, the gray and black line represent the trade-off curve obtained at steps 3 and 4, respectively. This implies that the dashed black line represents treatment plans where the dose rates are variable, but which were obtained using the plans corresponding to the gray dashed line (with all dose rates equal to 10MU/s) as a starting point for step 4.
	
	In addition to the trade-off curve, we show the fluence maps corresponding to one arc segment for each of the cases (Figures \ref{fig:flMapspr36} and \ref{fig:flMapsHN36}). This allows for comparison of the fluence maps obtained with individual arc segment optimization (first row), full arc optimization with optimized dose rates obtained at steps 4 and 3 (second and third row, respectively) and full arc optimization with constant maximum dose rates (fourth row). Note that the latter corresponds to the dashed gray line in the trade-off curves. Each of the columns corresponds to a different full arc delivery time ($TT$). The maps in the first row are optimized using the time allocated to this arc segment for the full arc delivery time corresponding to that column, namely for 6s, 9s, 15s, 18s and 18s for the prostate case, and 4s, 6s, 16s and 20s for the head and neck case. The corresponding ssdifs are indicated below each of the maps. Note that an increase in $TT$ may yield a lower fluence map resemblance for an individual arc segment. This is because the model optimizes row by row, not map by map, and a possible gain in an individual arc segment may be sacrificed for an overall gain.
	
	\subsection{Prostate case}
\begin{figure}[!h]
		\centering
		\includegraphics[width=0.6\textwidth]{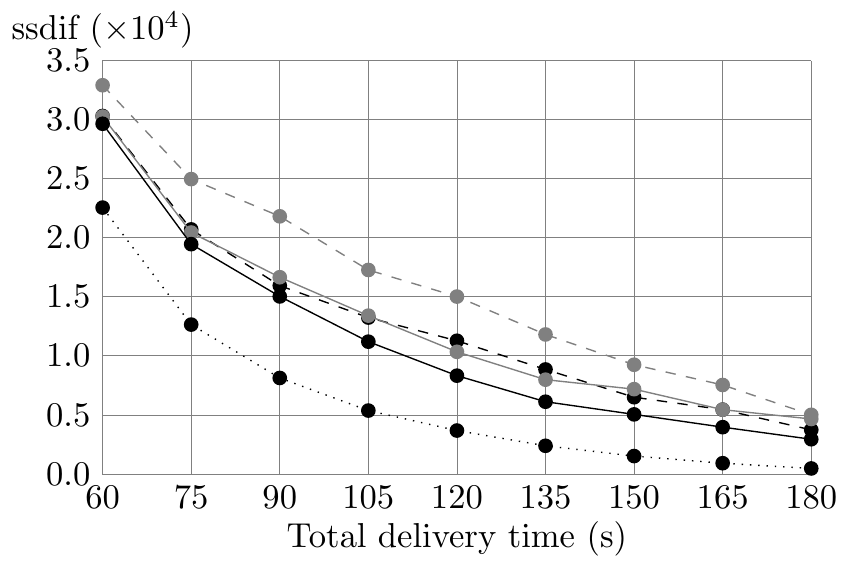}
		\caption{Trade-off between delivery time and ssdif for the prostate case. Results obtained at step 3 with optimized dose rates (solid) and  dose rates equal to 10 MU/s (dashed) are presented in gray. Improvements from these results obtained in step 4 are indicated in black. The lower bound based on individually optimized arc segments is shown by the dotted line.}\label{fig:prosTradeOff}
	\end{figure}

\begin{figure}[!htbp]
		\centering
		\includegraphics[width=0.8\textwidth]{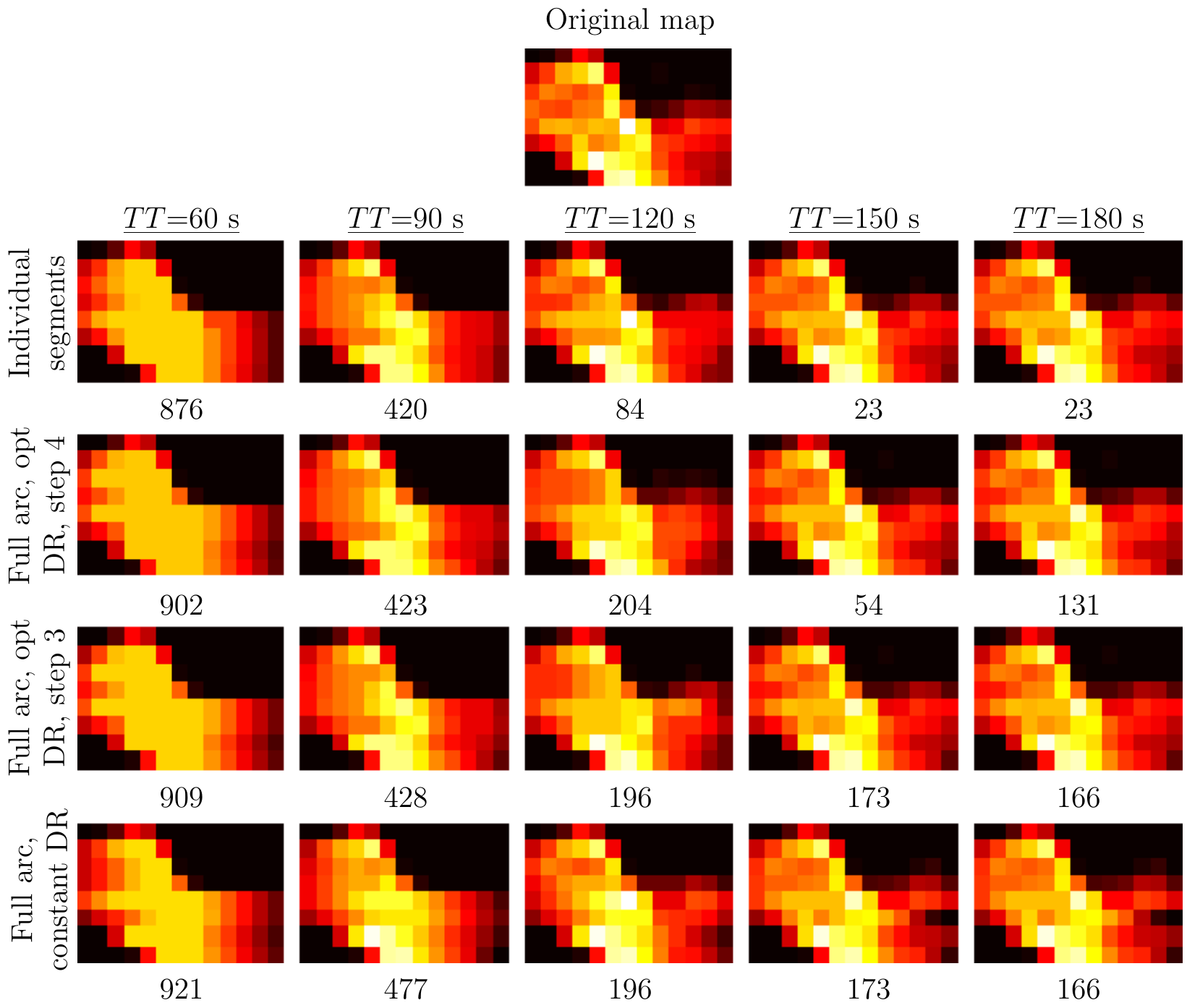}
		\caption{Fluence maps for arc segment 11 of the prostate case, optimized using a total arc delivery time ($TT$) of 60s, 90s, 120s, 150s and 180s. Fluence maps obtained with individual arc segment optimization are optimized with delivery times equal to the time allotted to this arc segment in the full arc treatment plans (6s, 9s, 15s, 18s and 18s, respectively). The corresponding ssdif is indicated below each map.}\label{fig:flMapspr36}
	\end{figure}

	The trade-off curve for the prostate case (Figure \ref{fig:prosTradeOff}) has a convex shape. This implies that for short delivery times, much can be gained by lengthening the treatment, whereas this is less so for treatments of average length. The same trend is visible in the fluence maps (Figure \ref{fig:flMapspr36}).
	
	The limitations of a constant maximum dose rate are evident from Figure \ref{fig:flMapspr36}: the gains from applying step 4 to the plans with constant dose rate are high. Furthermore, in order to achieve the same level of fluence map resemblance with constant dose rates as can be obtained with variable dose rates, the total delivery time needs to be increased, sometimes even by 45 seconds. Furthermore, allowing for a varying dose rate in step 3 gives better solutions in step 4.
	
\begin{figure}[!htbp]
		\centering
		\includegraphics[width=0.6\textwidth]{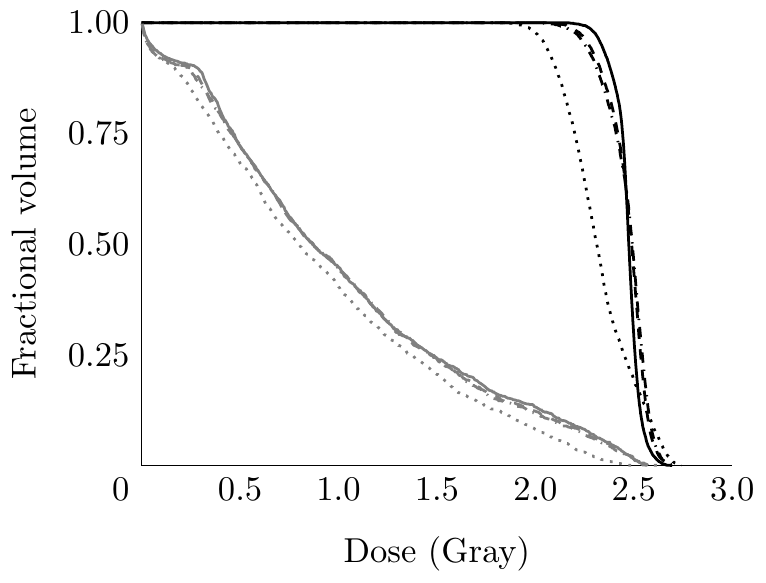}
		\caption{DVH plots for the PTV (black) and rectum (gray) for the plans obtained with $TT=60$ (dotted), $TT=150$ (dashdotted), $TT=180$ (dashed) and a sliding window approach (solid). Dose for a single fraction is shown. }\label{fig:DVH}
	\end{figure}

	The level of fluence map replication shown in Figures \ref{fig:prosTradeOff} and \ref{fig:flMapspr36} reflects the quality of DVH replication, as can be seen from Figure \ref{fig:DVH}. The DVHs for a total delivery time of 150s and 180s are slightly worse than those for the sliding window delivery (for which 240 seconds of delivery time are required), while the DVH for a delivery time of 60s shows a major deterioration.
	
\begin{figure}[!htbp]
		\centering
		\subfloat[]{\includegraphics[width=0.45\textwidth]{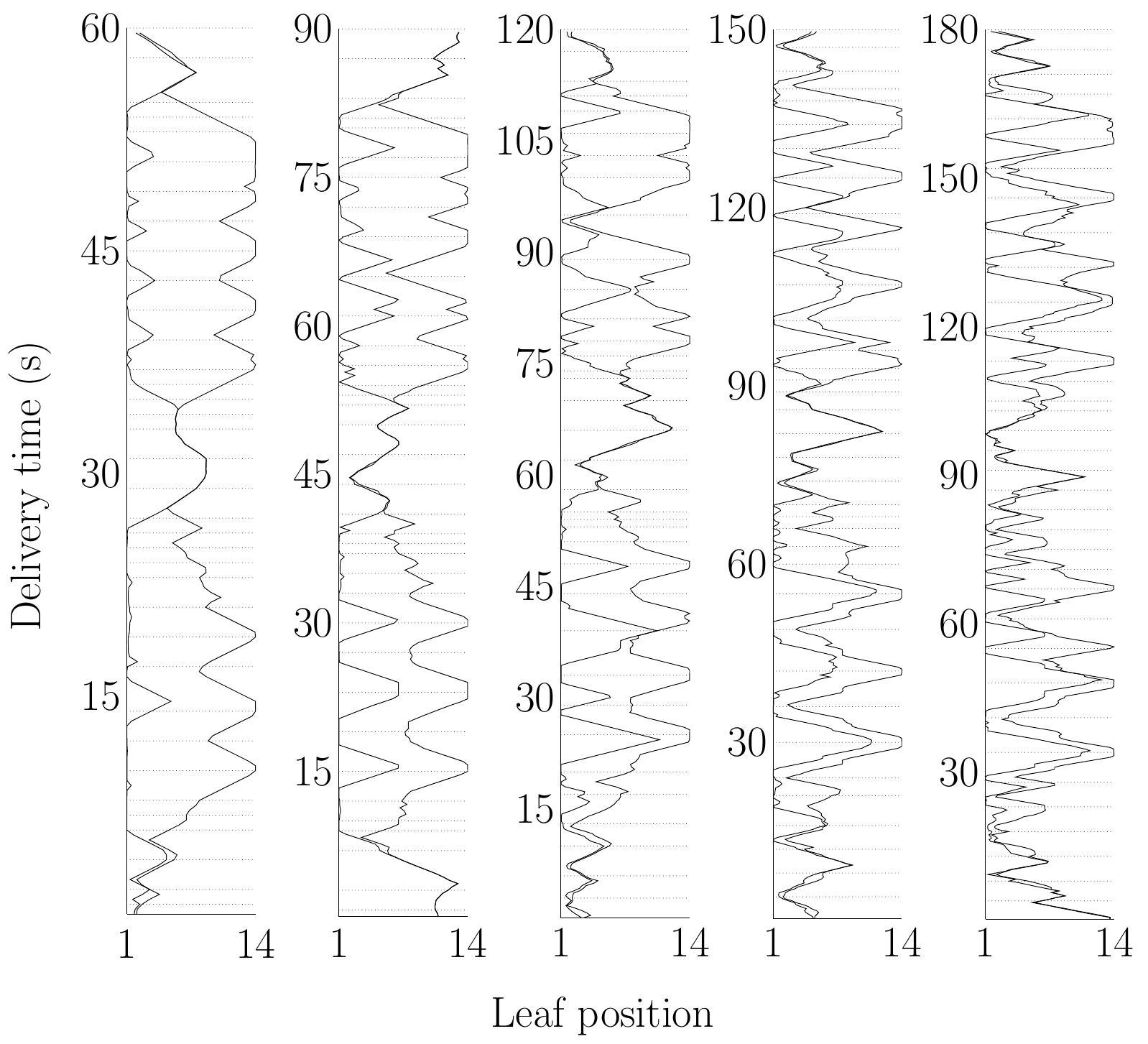}\label{fig:leafTrajDRoptConn}}
		\quad
		\subfloat[]{\includegraphics[width=0.45\textwidth]{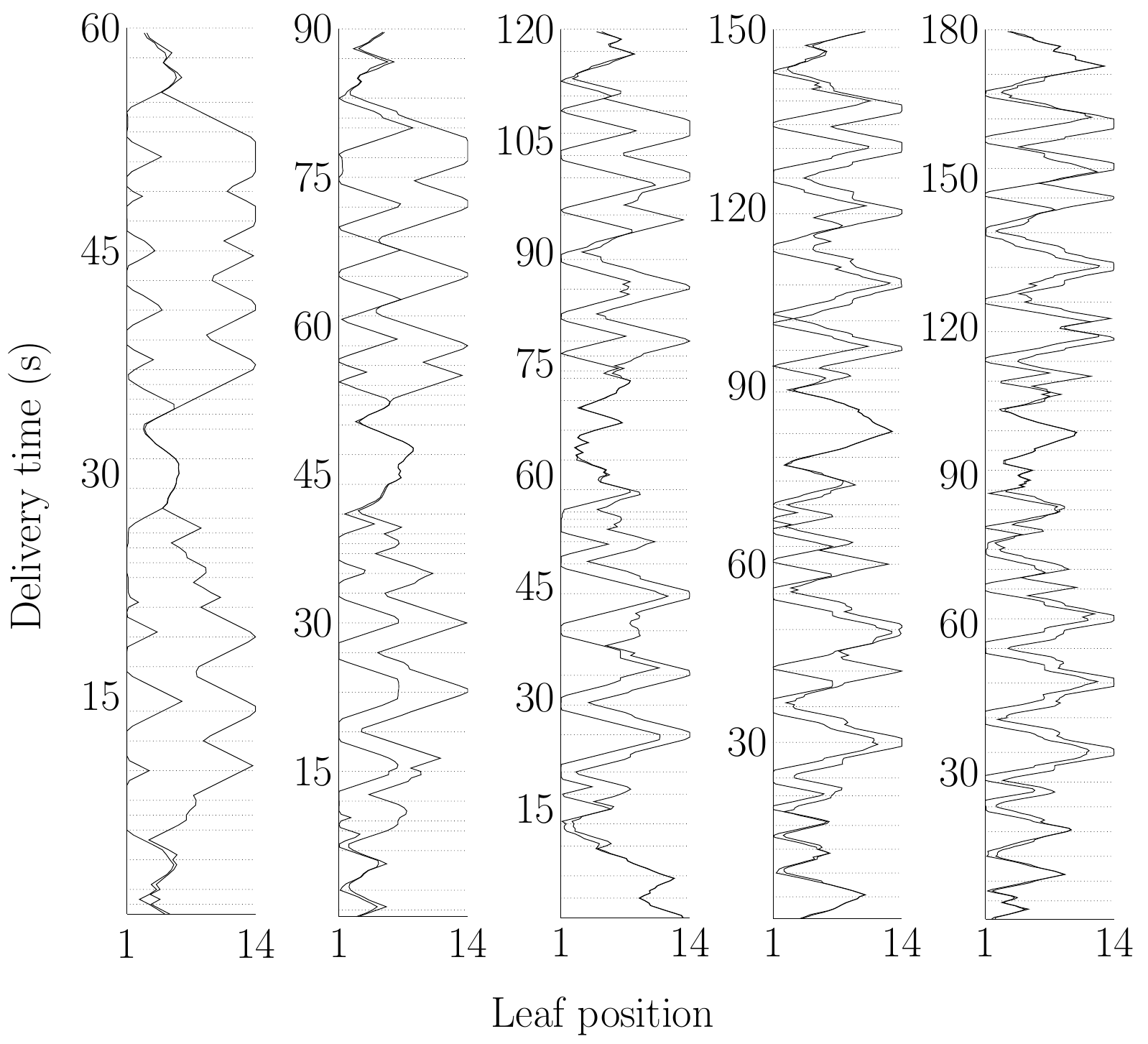}\label{fig:leafTrajDR10Conn}}
		\caption{The leaf trajectories for row 8 of the prostate case using variable (a) and constant maximum dose rates (b) for treatment durations from 1 (left-most plot) up to 3 minutes (right-most plot) with 30 second increments.}
	\end{figure}

	Figure \ref{fig:leafTrajDRoptConn} shows the leaf trajectories for treatment durations varying from 60 (left-most plot) up to 180 seconds (right-most plot) with 30 second increments. In the 1-minute treatment plan the leaves are generally either (almost) completely closed or they leave a large opening to deliver dose to many bixels at the same time. For the latter there is often a small close-in or open-out movement visible that results in the delivery of a peak-like fluence. As delivery time increases, the opening between the leaves generally shrinks and the trajectories become more similar to a sliding window approach. For a fixed dose rate, plans converge to the sliding window approach even faster (Figure \ref{fig:leafTrajDR10Conn}).
	
	Note that even though the leaves remain closed in some arc segments (see Figures \ref{fig:leafTrajDRoptConn} and \ref{fig:leafTrajDR10Conn}) as no fluence needs to be delivered here, a rather large amount of time is allocated to this arc segment. This is because another row of that fluence map requires a large amount of time for delivery.
	
	\subsection{Head and neck case}
\begin{figure}[!htbp]
		\centering
		\includegraphics{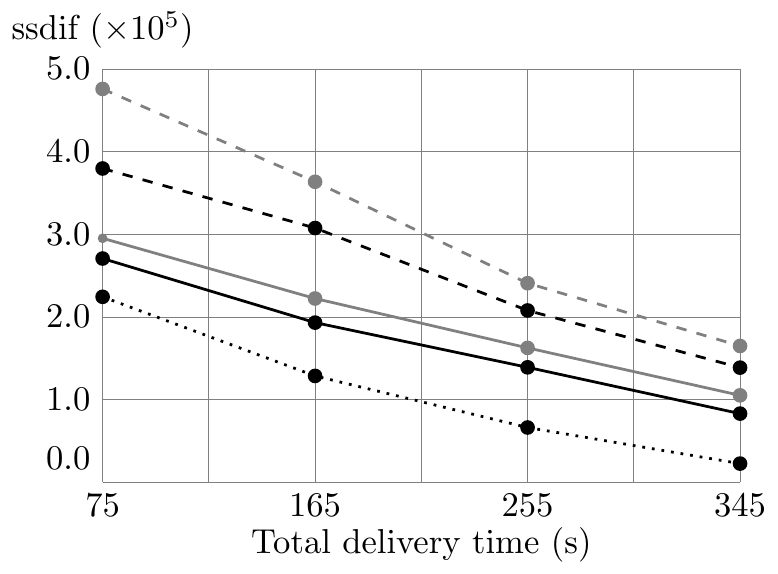}
		\caption{Trade-off between delivery time and ssdif for the head \& neck case. Results obtained at step 3 with optimized dose rates (solid) and  dose rates equal to 10 MU/s (dashed) are presented in gray. Improvements from these results obtained in step 4 are indicated in black. The lower bound based on individually optimized arc segments is shown by the dotted line.}\label{fig:HNTradeOff}
	\end{figure}

	While the trade-off curve for the prostate case is convex, it is nearly linear for the head and neck case (Figure \ref{fig:HNTradeOff}). The benefits from allowing for a varying dose rate are even more pronounced than for the prostate case: the delivery times can be reduced by up to 100s when using varying instead of constant dose rates, without reducing ssdif. Additionally, treatment plans with constant dose rates do not provide a good starting solution for step 4: the dashed black trade-off curve lies far above the trade-off curve obtained at step 3 when allowing for varying dose rates.
	
\begin{figure}[!htbp]
		\centering
		\includegraphics[width=0.75\textwidth]{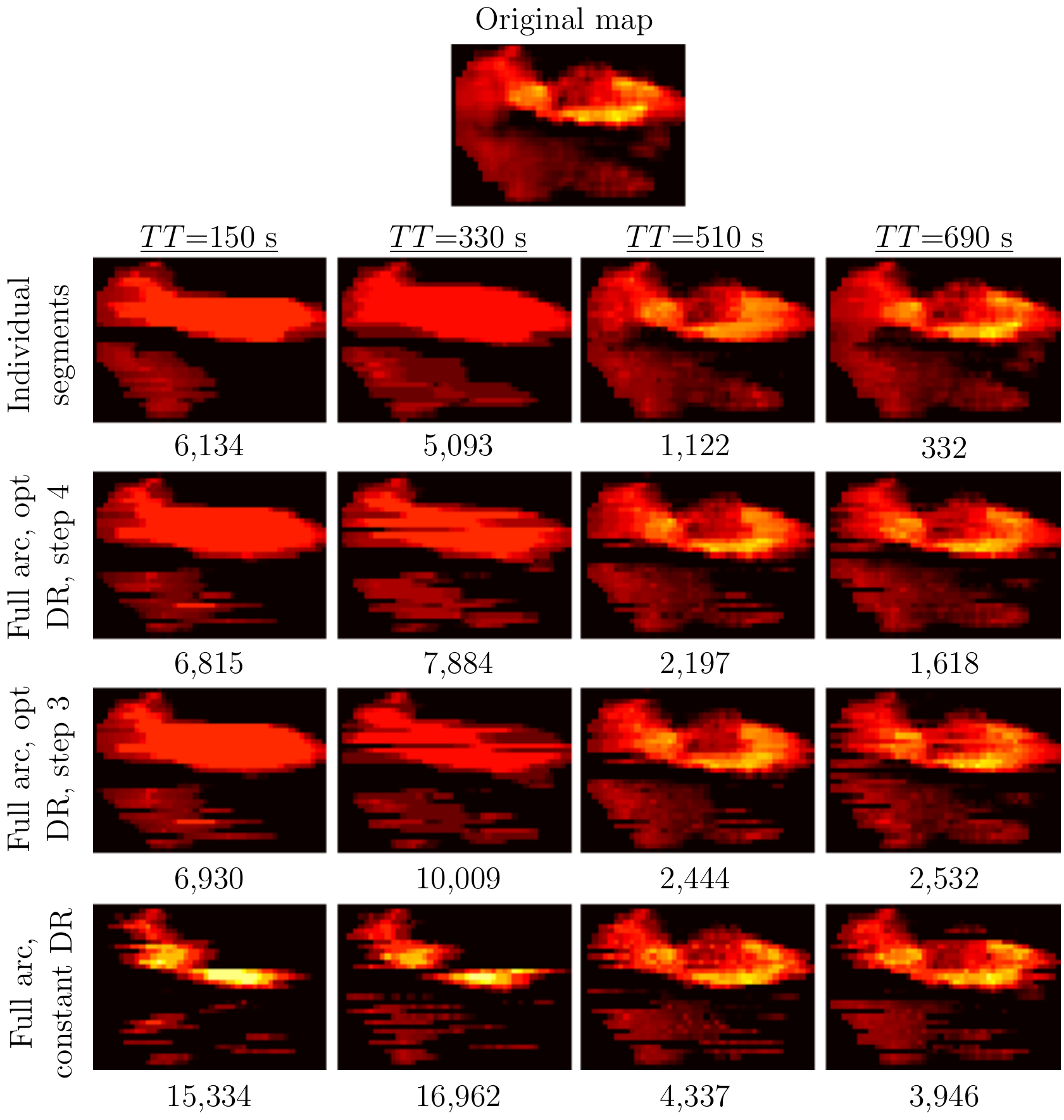}
		\caption{Fluence maps for arc segment 15 of the head and neck case, optimized using a total arc delivery time ($TT$) of 150s, 330s, 510s and 690s. Fluence maps obtained with individual arc segment optimization are optimized with delivery times equal to the time allotted to this arc segment in the full arc treatment plans (4s, 6s, 16s and 20s, respectively). The corresponding ssdif is indicated below each map.}\label{fig:flMapsHN36}	
	\end{figure}

\begin{figure}[!htbp]
		\centering
		\subfloat[]{\includegraphics[width=0.45\textwidth]{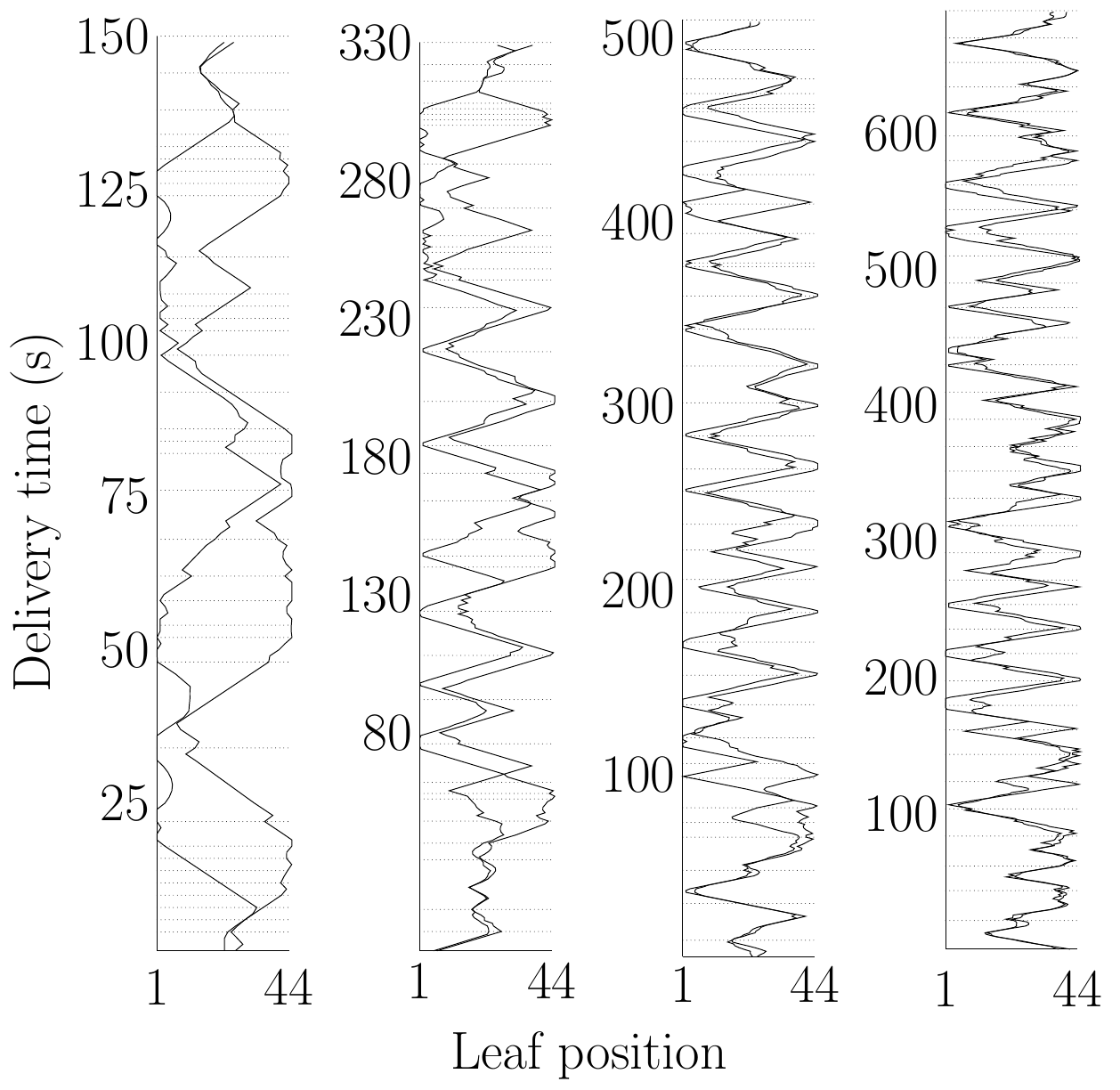}\label{fig:leafTrajDRoptHN}}
		\quad
		\subfloat[]{\includegraphics[width=0.45\textwidth]{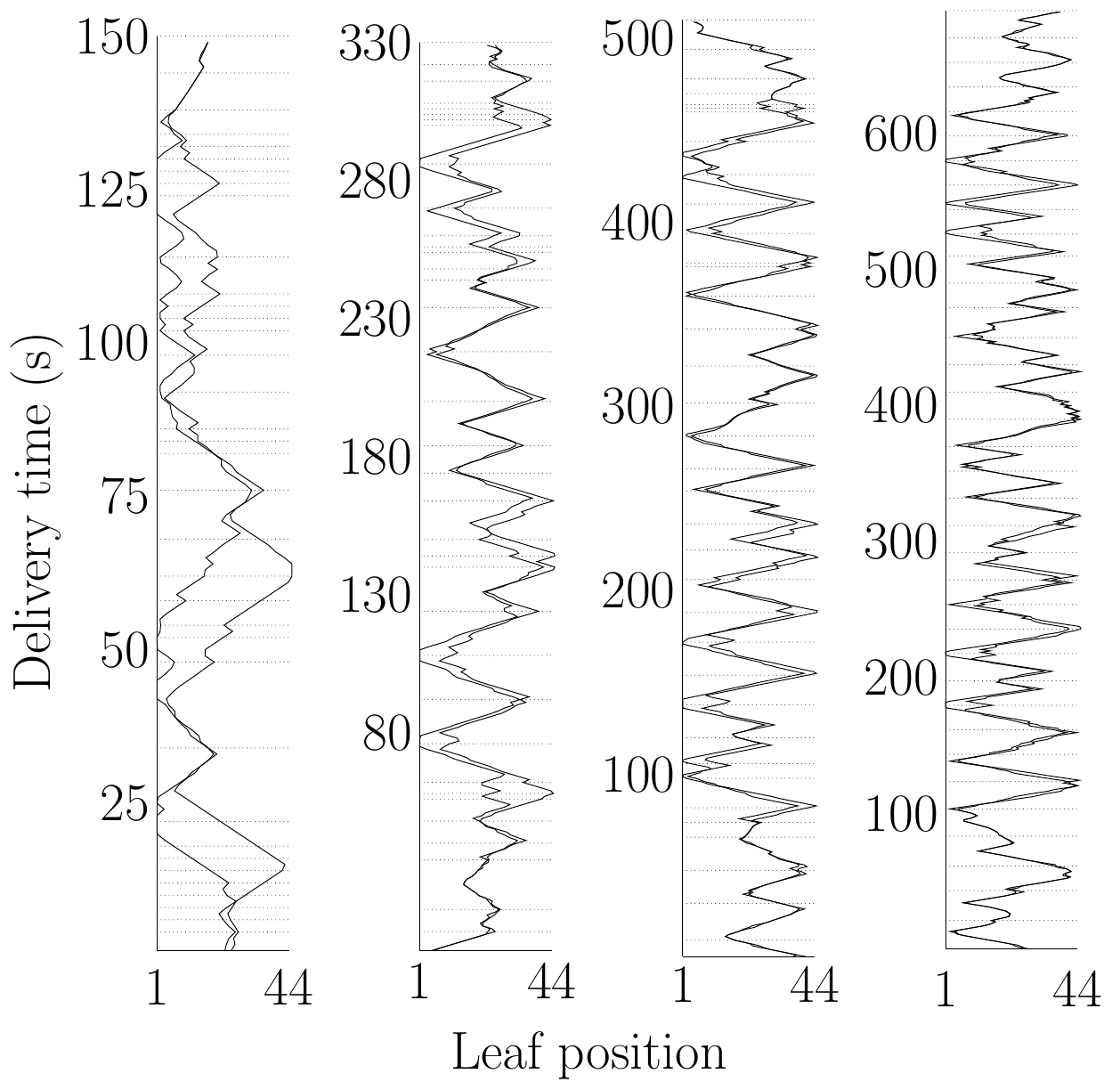}\label{fig:leafTrajDR10HN}}
		\caption{The leaf trajectories for row 22 of the head and neck case using variable (a) and constant maximum dose rates (b) for treatment durations from 75 (left-most plot) up to 345 seconds (right-most plot) with 90 second increments.}
	\end{figure}

	Figure \ref{fig:flMapsHN36} shows the fluence maps for arc segment 15 obtained with four different optimization approaches. The maps corresponding to a low delivery time ($TT=150$s and $TT=330$s, where respectively 4s and 6s are assigned to this map) show the different choices the optimization model makes when using a variable versus a constant maximum dose rate. The short delivery times limit the number of bixels a leaf can traverse. As a result, the leaves can either be placed close to the boundaries of the field to deliver fluence to a large part of the row, or they can be located close to one another, which implies that they block many of the bixels. In the case of a variable dose rate, the model chooses the first option and turns the dose rate down in order to avoid an excessively high amount of fluence to the bixels that are exposed. However, when fixing the dose rate to its maximum, one cannot avoid this excessively high amount of fluence, so the model is forced to place the leaves closer to each other, thus blocking out a large part of the row. This also becomes apparent from Figures \ref{fig:leafTrajDRoptHN} and \ref{fig:leafTrajDR10HN}, showing the leaf trajectories for row 20 using optimized and constant maximum dose rates, respectively.
	
	\section{Discussion}
	
	In this article we have developed a method to construct the delivery time versus fluence map fidelity trade-off curve for VMAT. The core of our method is an optimization model that optimizes the leaf trajectories and dose rates for a given delivery time such that the optimal fluence maps are replicated as well as possible. Solving this optimization model for a sequence of delivery times gives the complete trade-off curve, and allows the planner to balance delivery time and fluence map resemblance. The method is an extension of the single map trade-off curve method from \cite{dm1}.
	
	The trade-off curves for the prostate case presented in this article show that the gain in fluence map resemblance from an additional second of delivery time decreases as the total delivery time increases. More importantly, using a shorter delivery time than what is required for the sliding window approach yields only a minor reduction in plan quality. A further reduction may result in too low fluence map resemblance. For the head and neck case, there seems to be a more linear relation between delivery time and fluence map resemblance. Construction of the trade-off curve allows the planner to obtain these insights and choose the preferred balance between the delivery time and the level of fluence map replication.
	
	While the majority of the current literature assumes a constant dose rate, we include optimization of the dose rate scheme as well. Comparing the trade-off curves of plans with a varying dose rate to those with a constant maximum dose rate shows that the same level of plan quality can be achieved with shorter delivery times when using variable dose rates. A similar conclusion was found by \cite{Palma:2008}. The use of variable dose rates is already required in step 3 of the algorithm: a better starting solution for step 4 results in a better final solution.
	
	Several other works have looked into the conversion of fluence maps to a deliverable treatment plan as well. Some of these use a sliding window delivery since this allows the maps to be replicated to high accuracy \citep{vmerge,cortdoi}, though it inevitably results in a high treatment time as the leaves have to traverse the whole field for each fluence map. This has led to the development of several arc sequencing algorithms that allow for other types of leaf trajectories as well. \cite{Shepard:2007} minimize the sum of absolute differences between the ideal and planned fluence maps using simulated annealing. The authors optimize the aperture shapes and dose rates at every 10$^{\circ}$ (which can be chosen to be more dense). Their objective is to minimize the sum of absolute differences between the ideal and planned fluence maps. The optimization problem is similar to ours as both methods directly aim to minimize the difference between ideal and planned fluence maps, though the model is solved using a different solution method (simulated annealing versus an interior point method). \cite{Bokrantz:2012} also use optimized fluence maps, but do not consider directly replicating these. Instead, their goal is to find a treatment plan that minimizes the difference between the DVHs corresponding to the optimized and the planned maps. A comparison of the performance of these methods is only possible when using the same set of patients.
	
	\cite{Unkelbach:2015} mention several disadvantages of treatment planning via fluence maps compared to DAO, which are mostly overcome by our method. First, the formulation of a DAO model exactly reflects the DICOM specifications and thus always yields a deliverable treatment plan. In our model, we use the same decision variables as are generally used in DAO approaches, and the resulting treatment plan can thus be described according to DICOM specifications. Second, \cite{Unkelbach:2015} state that a highly accurate replication of fluence maps is expected to yield inefficient treatment delivery. This is exactly the issue we are addressing here by constructing the delivery time versus replication accuracy trade-off curve. A third and final drawback is that the distribution of apertures over an arc sector inherent to arc sequencing generally results in a dose degradation. This drawback does hold for our method. As shown in \cite{vmerge}, the degradation is minimal when arc sectors of 10$^{\circ}$ or less are used. We developed our work to bridge the gap between the coarse, large field, and fast delivery solutions typical of direct aperture approaches and the smaller field, higher monitor unit, and more time consuming sliding window approaches.
	
	This work is the first to assess the trade-off between delivery time and fluence map resemblance for arc sequencing, but it is not the first to assess the general question of the tradeoff between delivery time and plan quality in VMAT. \cite{vmerge} explore quality versus time for by optimizing fluence maps at a 2$^{\circ}$ angular spacing and then successively merging fluence maps based on their similarity. Each merge decreases treatment time but also decreases plan quality. In this approach, the fluence maps are replicated perfectly using sliding window delivery. This prompts two further ideas to achieve delivery time versus plan quality trade offs: either by adding a smoothing term to the original fluence map optimization--an approach that has recently been investigated by \cite{gaddyANDpapp2016}--or by speeding up the delivery of the fluence maps by departing from strict sliding window, as we do herein. The optimal way to combine all three of these approaches--merging, smoothing, and approximate delivery--remains to be worked out.

	The main limitation of our method is the size of the full arc optimization model, which requires us to use a heuristic approach. This has several consequences. First, due to the highly non-convex nature of the model, we have no guarantee on the optimality of our solution. However, as the objective function value plateaus as the number of starting points for which the model is solved increases, we believe that the optimal ssdif found in step 4 lies close to the true optimal objective value of this model. Second, a large amount of time is required to solve the model, which makes it unsuitable for implementation in the clinic in its current form. The size of the problem increases with the size of the fluence maps and the total delivery time. Therefore, future research should aim at reducing the solution times. For this, one can take advantage of the highly parallelizable nature of the method. Furthermore, for some rows there are several consecutive maps where the fluence is (close to) zero, which may allow us to ignore those segments in the first optimization, and optimize their leaf trajectories in a post optimization. Lastly, heuristics to find better starting solutions, or other techniques to sample the global solution space, will reduce the number of local searches that need to be performed.
	
	\section{Conclusion}
	The trade-off between delivery time and fluence map replication accuracy lies at the heart of VMAT treatment planning. In this work we have presented a method to construct the trade-off curve between these two conflicting objectives by optimizing treatment plans for various delivery times. Future research is needed to speed up the optimization process before it can be clinically implemented. Once this becomes possible, the treatment planner can use the trade-off curve to select a plan that balances delivery time and plan quality.

%\section*{References}
\bibliographystyle{plainnatnourl}
\bibliography{dm2}

\begin{thebibliography}{14}
\providecommand{\natexlab}[1]{#1}
\providecommand{\url}[1]{\texttt{#1}}
\expandafter\ifx\csname urlstyle\endcsname\relax
  \providecommand{\doi}[1]{doi: #1}\else
  \providecommand{\doi}{doi: \begingroup \urlstyle{rm}\Url}\fi

\bibitem[Bokrantz(2012)]{Bokrantz:2012}
R.~Bokrantz.
\newblock Multicriteria optimization for volumetric-modulated arc therapy by
  decomposition into a fluence-based relaxation and a segment weight-based
  restriction.
\newblock \emph{Med Phys}, 39\penalty0 (11):\penalty0 6712--25, 2012.

\bibitem[Chen et~al.(2011)Chen, Luan, and Wang]{coupled}
D.~Chen, S.~Luan, and C.~Wang.
\newblock Coupled path planning, region optimization, and applications in
  intensity-modulated radiation therapy.
\newblock \emph{Algorithmica}, 60\penalty0 (1):\penalty0 152--174, 2011.

\bibitem[Chen et~al.(2012)Chen, Unkelbach, Trofimov, Madden, Kooy, Bortfeld,
  and Craft]{chen2012including}
W~Chen, J~Unkelbach, A~Trofimov, T~Madden, H~Kooy, T~R Bortfeld, and D~L Craft.
\newblock Including robustness in multi-criteria opimization for intensity
  modulated proton therapy.
\newblock \emph{Phys Med Biol}, 57\penalty0 (3):\penalty0 591--608, 2012.

\bibitem[Craft and Balvert()]{dm1}
D.~Craft and M.~Balvert.
\newblock Fast approximate delivery of fluence maps: the single map case.

\bibitem[Craft et~al.(2012)Craft, McQuaid, Wala, Chen, Salari, and
  Bortfeld]{vmerge}
D.~Craft, D.~McQuaid, J.~Wala, W.~Chen, E.~Salari, and T.~Bortfeld.
\newblock Multicriteria {VMAT} optimization.
\newblock \emph{Med Phys}, 39\penalty0 (2):\penalty0 686, 2012.

\bibitem[Craft et~al.(2014)Craft, Bangert, Long, Papp, and Unkelbach]{cortdoi}
D.~Craft, M.~Bangert, T.~Long, D.~Papp, and J.~Unkelbach.
\newblock The {CORT} dataset.
\newblock \emph{GigaSci Database}, 3\penalty0 (1):\penalty0 1, 2014.

\bibitem[Gaddy and Papp(2016)]{gaddyANDpapp2016}
M.~R. Gaddy and D.~Papp.
\newblock Technical note: Improving the {VMERGE} treatment planning algorithm
  for rotational therapy.
\newblock \emph{Med Phys}, 43\penalty0 (7):\penalty0 4093--7, 2016.

\bibitem[Kierkels et~al.(2015)Kierkels, Visser, Bijl, Langendijk, Van~'t Veld,
  Steenbakkers, and Korevaar]{Kierkels:2015}
R.~G.~J. Kierkels, R.~Visser, H.~P. Bijl, J.~A. Langendijk, A.~Van~'t Veld,
  R.~J. H.~M. Steenbakkers, and E.~W. Korevaar.
\newblock Multicriteria optimization enables less experienced planners to
  efficiently produce high quality treatment plans in head and neck cancer
  radiotherapy.
\newblock \emph{Radiat Oncol}, 10\penalty0 (1):\penalty0 1, 2015.

\bibitem[Luan et~al.(2008)Luan, Wang, Cao, Chen, Shepard, and
  Cedric]{Luan:2008}
S.~Luan, C.~Wang, D.~Cao, D.~Chen, D.~Shepard, and X.~Cedric.
\newblock Leaf-sequencing for intensity-modulated arc therapy using graph
  algorithms.
\newblock \emph{Med Phys}, 35\penalty0 (1):\penalty0 61--69, 2008.

\bibitem[Palma et~al.(2008)Palma, Vollans, James, Nakano, Moiseenko, Schaffer,
  McKenzie, Morris, and Otto]{Palma:2008}
D.~Palma, E.~Vollans, K.~James, S.~Nakano, V.~Moiseenko, R.~Schaffer,
  M.~McKenzie, J.~Morris, and K.~Otto.
\newblock Volumetric modulated arc therapy for delivery of prostate
  radiotherapy: comparison with intensity-modulated radiotherapy and
  three-dimensional conformal radiotherapy.
\newblock \emph{Int J Rad Oncol Biol Phys}, 72\penalty0 (4):\penalty0
  996--1001, 2008.

\bibitem[Shepard et~al.(2007)Shepard, Cao, Afghan, and Earl]{Shepard:2007}
D.M. Shepard, D.~Cao, M.K.N Afghan, and M.A. Earl.
\newblock An arc-sequencing algorithm for intensity modulated arc therapy.
\newblock \emph{Med Phys}, 34\penalty0 (2):\penalty0 464--70, 2007.

\bibitem[Stein et~al.(1994)Stein, Bortfeld, D\"orschel, and Schlegel]{Stein:94}
J.~Stein, T.~Bortfeld, B.~D\"orschel, and W.~Schlegel.
\newblock Dynamic x-ray compensation for conformal radiotherapy by means of
  multi-leaf collimation.
\newblock \emph{Radiother Oncol}, 32\penalty0 (2):\penalty0 163--173, 1994.

\bibitem[Teoh et~al.(2011)Teoh, Clark, Wood, Whitaker, and Nisbet]{Teoh:2011}
M~Teoh, C.~Clark, K.~Wood, S.~Whitaker, and A.~Nisbet.
\newblock Volumetric modulated arc therapy: a review of current literature and
  clinical use in practice.
\newblock \emph{Brit J Radiol}, 84\penalty0 (1007):\penalty0 967--96, 2011.

\bibitem[Unkelbach et~al.(2015)Unkelbach, Bortfeld, Craft, Alber, Bangert,
  Bokrantz, Chen, Li, Xing, Men, Nill, Papp, Romeijn, and
  Salari]{Unkelbach:2015}
J.~Unkelbach, T.~Bortfeld, D.~Craft, M.~Alber, M.~Bangert, R.~Bokrantz,
  D.~Chen, R.~Li, L.~Xing, C.~Men, S.~Nill, D.~Papp, E.~Romeijn, and E.~Salari.
\newblock Optimization approaches to volumetric modulated arc therapy planning.
\newblock \emph{Med Phys}, 42\penalty0 (3):\penalty0 1367, 2015.

\end{thebibliography}

\end{document}